\RequirePackage[2020-02-02]{latexrelease}
\documentclass[aps,prl,amsfonts,preprint]{revtex4}

\usepackage{dcolumn}
\usepackage{amsmath}
\usepackage{bm}
\usepackage{graphicx}

\usepackage{graphics}
\usepackage{longtable}
\usepackage{color}
\usepackage{epsfig}
\usepackage[normalem]{ulem}

\begin{document}
\def\Tc{$T_{\mathrm C}$}
\def\TN{$T_{\mathrm N\'eel}$}
\def\Ef{$E_{\mathrm F}$}
\def\BiTe{Bi$_2$Te$_3$}
\def\BiSe{Bi$_2$Se$_3$}
\def\SbTe{Sb$_2$Te$_3$}
\def\BiSbTe{(Bi, Sb)$_2$Te$_3$}
\def\SbVTe{(Sb$_{1-x}$V$_x$)$_2$Te$_3$}
\def\BiMnTe{(Bi$_{1-x}$Mn$_x$)$_2$Te$_3$}
\def\BiMnSe{(Bi$_{1-x}$Mn$_x$)$_2$Se$_3$}
\def\MnSbSbTe4{MnSb$_2$Te$_4$}
\def\MnBiBiTe4{MnBi$_2$Te$_4$}
\def\MnBiBiSe4{MnBi$_2$Se$_4$}
\def\HgMnTe{Hg$_{1-x}$Mn$_x$Te}
\def\Gbar{$\overline{\Gamma}$}
\def\MnBi{Mn$_{\rm Bi}$}
\def\muB{$\mu_{\rm B}$}
\def\dIdV{${\rm d}I/{\rm d}V$}
\def\PK{\textcolor{blue}}
\def\MM{\textcolor{red}}
\def\OR{\textcolor{black}}
\def\bl{\textcolor{black}}
\def\GS{\textcolor{magenta}}
%\pagestyle{fancy}
%\rhead{\includegraphics[width=2.5cm]{vch-logo.png}}

\title{Probing the magnetic band gap of the ferromagnetic topological insulator \MnSbSbTe4}

\author{P. K\"uppers$^{1}$, J. Zenner$^{1}$, S. Wimmer$^{2}$, G. Springholz$^{2}$, O. Rader$^{3}$, M. Liebmann$^1$, M. Morgenstern$^{1*}$}  % end authors

\affiliation{$^1$II. Institute of Physics B and JARA-FIT, RWTH Aachen University, 52074 Aachen, Germany}

\affiliation{$^2$Institut f\"ur Halbleiter- und Festk\"orperphysik, Johannes Kepler Universit\"at,
			Altenberger Stra\ss e 69, 4040 Linz, Austria}
			
\affiliation{$^3$Helmholtz-Zentrum Berlin f\"ur Materialien und Energie,
			 Albert-Einstein-Stra\ss e 15, 12489 Berlin, Germany}

\date{\today}

\begin{abstract}
{\bf {Mn-rich \MnSbSbTe4\ is a ferromagnetic topological insulator with yet the highest Curie temperature $T_{\rm C}=45-50$\,K. It  exhibits a magnetic gap at the Dirac point of the topological surface state that disappears above $T_{\rm C}$. By scanning tunneling spectroscopy, we probe this gap at different magnetic fields and temperatures. We firstly reveal that the gap size shrinks, when an in-plane magnetic field of up to $B_{\parallel}= 3$\,T is applied, but does not close completely as the magnetization is only partially rotated in-plane. This corroborates the magnetic origin of the gap and the complex magnetic structure. In addition, we demonstrate significant spatiotemporal fluctuations of the gap size at temperatures as low as $T_{\rm C}/2$}, above which the remanent magnetization indeed decays. This temperature is close to the antiferromagnetic transition temperature observed for bulk-type single crystals of \MnSbSbTe4, highlighting the important role of competing magnetic orders in the formation of the favorable ferromagnetic topological insulator. Our study, thus,  provides crucial insights into the complex magnetic gap opening of topological insulators that is decisive for quantum anomalous Hall devices.}

\end{abstract}
  \maketitle
\noindent {{Corresponding author: } 
M.~Morgenstern, email: mmorgens@physik.rwth-aachen.de %O. Rader, email: rader@helmholtz-berlin.de, G. Springholz, email: gunther.springholz@jku.at.
} % end rouge 

%\noindent {$^+$ These authors contributed equally to the present work.} % end rouge 
\newpage
\vspace{2cm}
%______________________________________________________________________________
% Text: Please use section headings and subheadings as specified below. For communications, all section headings apart from Experimental Section should be removed
% Please make the first reference to a display item bold: \textbf{Figure 1}
% Do not abbreviate Figure, Equation, etc.; display items are always singular, i.e., Figure 1 and 2.
% Equations are always singular, i.e., Equation 1 and 2, and should be inserted using the {equation} environment, not as graphics
% Please do not use footnotes in the text, additional information can be added to the Reference list.

The quantum anomalous Hall effect (QAHE)  offers quantized conductance and lossless transport without the need for an external magnetic field $B$ \cite{Onoda03,CXLiu08,YuScience10}. This implies applications, e.g., in metrology \cite{FoxPRB18}.
Combining topological insulators (TIs) with transition metal atoms \cite{CXLiu08,YuScience10,Qiao10,Tokura19} led to %non-volatile magnetism and to  
the QAHE in Cr- and V- doped \BiSbTe\   \cite{Chang13,CheckelskyNP14,KouPRL14,BestwickPRL15,Kandala2015} with exact quantization down to a precision of $10^{-8}$ \cite{ChangCZNM15,GrauerPRB15,FoxPRB18,GoetzAPL18,Okazaki2021}. Fundamentally, the effect is based on the out-of-plane magnetization of ferromagnetically coupled transition metal atoms that opens a gap within the topological surface state. The gap features a non-trivial Chern number via the Berry curvature of the topological Dirac cone   \cite{Mong2010,Zhang2019,CXLiu08,YuScience10,Qiao10,Tokura19}. Surprisingly, however, the experimental temperatures  $T$ required for observing the QAHE ({$T<300$}\,mK \cite{Chang13,GrauerPRB15}) were initially two orders of magnitude lower than the ferromagnetic  transition temperature $T_{\rm C}\approx 20$\,K  \cite{Zhou2005,Figueroa2020,CheckelskyNP14}.
%If the temperature of the QAHE could be raised, applications such as chiral interconnects \cite{interconnects}, edge state spintronics \cite{Yasuda17,Mahoney2017} and metrological standards \cite{FoxPRB18,GoetzAPL18}  become realistic. 
%
This has been improved by modulation doping, i.e., by concentrating the magnetic dopants in distinct layers of the topological insulator and, thus, reducing the disorder \cite{MogiAPL15,Xiao2018}. The QAHE then survives up to $T=2$\,K, still much lower than \Tc\ \cite{MogiAPL15,Xiao2018}, which is likely due to fluctuations of the Dirac point energy caused by electrostatic disorder \cite{Chong2020,Fijalkowski2021}. 

\begin{figure}
\includegraphics*[scale=0.25]{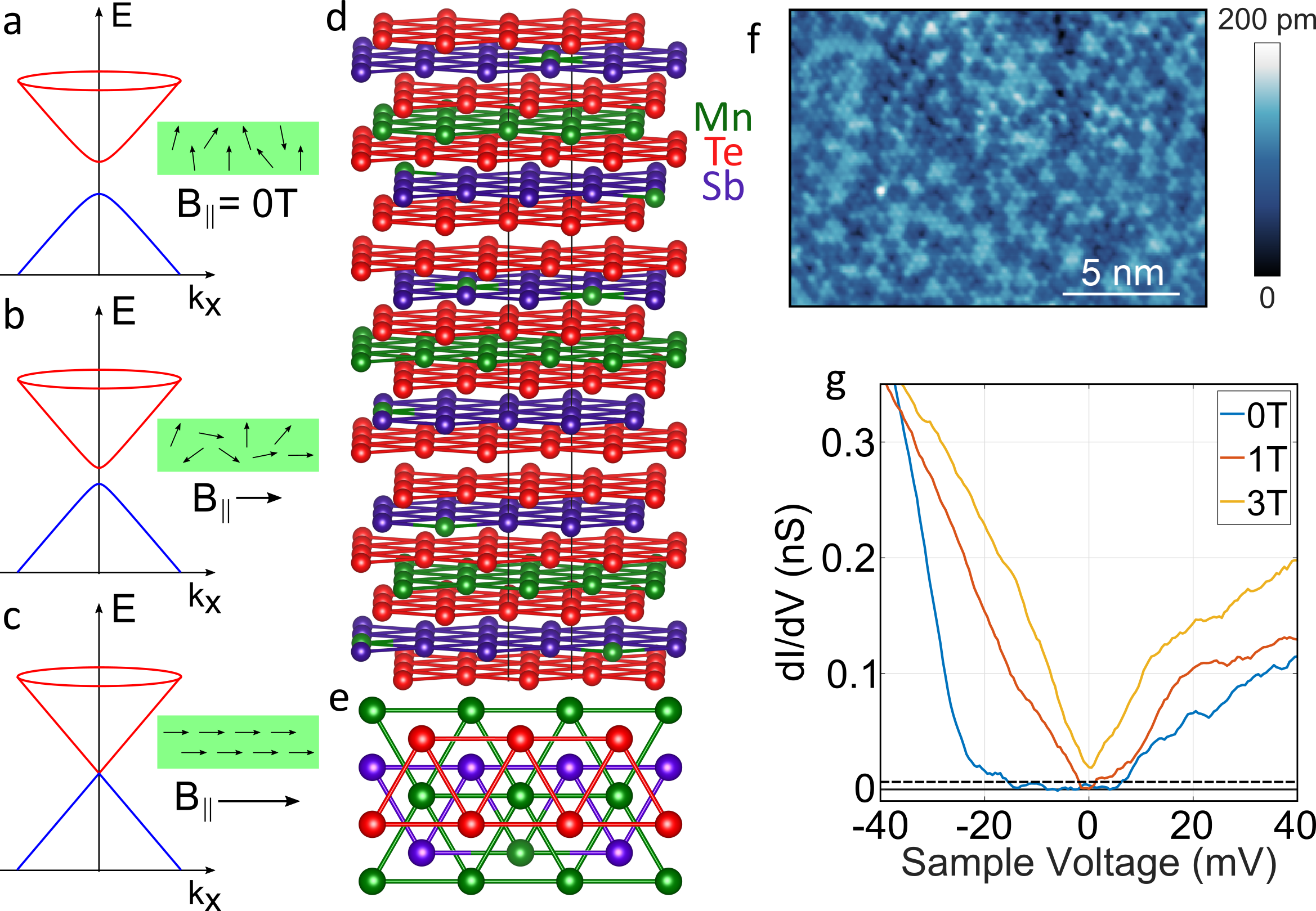}
\vspace{0cm}
\caption{{\bf} (a) --(c) Evolution of the topological surface state of ferromagnetic \MnSbSbTe4\ as a function of an applied $B_\parallel$ (left) that increasingly aligns the spins  of the Mn atoms (arrows in the green rectangle) towards the in-plane direction. (d) Side view of the stick and ball model of Mn-rich \MnSbSbTe4\ consisting of septuple layers Te-Sb-Te-Mn-Te-Sb-Te with van-der-Waals gaps between adjacent Te layers (red). Excess Mn atoms (green) are additionally distributed in the Sb layers (violet). The vertical lines are guides to the eye. (e) Top view of the first three layers. (f) Atomically resolved STM image of Mn-rich \MnSbSbTe4, $V=500$\,mV, $I=200$ pA. %Inset: Higher resolution image. 
(g) $dI/dV (V)$ curves recorded at the same position, but at different $B_\parallel$ as marked, $V_{\rm stab}=-100$\,mV, $I_{\rm stab}=400$\,pA. The dashed line marks the threshold for the gap determination, i.e. $dI/dV$ values above the line are not regarded as gap. The band gap closes with increasing $B_\parallel$. (f--g) $T=6$\,K. }
\label{Fig1}
\end{figure}

In contrast, intrinsic magnetic topological insulators (MTIs) enabled the observation of the QAHE up to about \Tc\  \cite{DengNatPhys2020}. 
Such MTIs consist of MnX$_2$Te$_4$ septuple layers (X=Bi or Sb)   with  Te-X-Te-Mn-Te-X-Te stacking \cite{Rienks,HagmannNJP17,Wimmer2021} (Fig.~\ref{Fig1}d). Hence, one extra layer of MnTe is inserted into each quintuple layer X$_2$Te$_3$ of conventional topological insulators \cite{YuScience10, YeNatComm2015, LiMPRL15}. This aims at reducing the disorder compared to randomly substituting X by Cr or V. Accordingly, even for Mn$_{x}$X$_{2}$Te$_{3+x}$ with $x<1$, the Mn containing septuple layers are regularly distributed between unperturbed quintuple layers \cite{Hu2020,Rienks}. For such Mn$_{x}$Bi$_{2}$Te$_{3+x}$, the QAHE indeed appeared up to $T \simeq 7$\,K (precison $\approx 10$\%), i.e., very close to the magnetic transition at $T=10$\,K \cite{DengNatPhys2020}.  
%Intriguingly, vacancies had to be induced to counteract the usual n-type doping of \BiTe. 
On the other hand, for Mn$_{x}$Bi$_{2}$Te$_{3+x}$, the appearance of a magnetically induced gap in the topological surface state  \cite{EremeevJAC17,McQueeneyPRM19,McQueeneyPRB19,OtrokovAFMTI18,ChenPRM20,Sekine2021} has remained experimentally controversial  \cite{Rienks,Yuan2020,OtrokovAFMTI18,PRX1,PRX3}, which is likely due to its strong sensitivity on sample preparation details as well as due to the predominantly antiferromagnetic coupling of the Mn layers \cite{Shikin2021,Garnica2022}.  In contrast, we have recently revealed that \MnSbSbTe4\ with slight %$\sim 7$\,\%
Mn excess is a ferromagnetic topological insulator (TI) with record high $T_{\rm C}=45-50$\,K and a pronounced band gap at the Fermi level $E_{\rm F}$ that opens precisely at and below \Tc\ \cite{Wimmer2021}. Moreover, the  band gap size $\Delta(T)$ was found to be proportional to the magnetization $M_\perp(T)$ perpendicular to the surface  as theoretically expected \cite{Rosenberg12,Henk2012}. The behaviour of  $M_\perp$ is, however rather complex \cite{Wimmer2021,Ge2021,Lai2021} as it reduces linearly with $T$ between
\Tc\ /2 and \Tc\ and exhibits a non-saturating magnetic moment of $\approx 1.5\,\mu_{\rm B}$ per Mn atom at $B=5$\,T \cite{Wimmer2021}. In fact, ferromagnetic \MnSbSbTe4\ with slightly different stoichiometry displays a saturation of the magnetization ($5\,\mu_{\rm B}$ per Mn atom) at $B=60$\,T only \cite{Lai2021}. In addition, nominally stoichiometric \MnSbSbTe4\ shows both ferromagnetism \cite{MurakamiPRB19,Ge2021,liu2021} and antiferromagnetism \cite{McQueeneyPRB19,ChenPRM20,liu2021} depending on the synthesis conditions. 
This complex magnetic behaviour is most likely related to the partial Mn-Sb exchange within the septuple layers in the few \% regime, which strongly depends on the growth conditions %as discovered by scattering methods  
\cite{liu2021,Riberolles2021,li2021,HuarXiv2020,Wimmer2021,Lai2021}. On the other hand, it is exactly this Mn-Sb exchange within the Sb layers that is responsible for the favorable combination of TI and ferromagnetic properties in Mn-rich \MnSbSbTe4\ \cite{Wimmer2021,li2021}.

%A ferromagnetic AHE has also been observed {for  systems} with either a larger amount of quintuple layers \cite{Vidal2019,HuSciAdv20,chen2020} or via alloying of Sb and Bi in Mn(Bi$_{2-x}$Sb$_x$)$_2$Te$_4$ \cite{McQueeneyPRB19,ChenPRM20,Shi2020} or both \cite{HuarXiv2020,huan2021}.

Here, we probe the change of the ferromagnetic gap in  epitaxial \MnSbSbTe4\ 
by  scanning tunneling spectroscopy (STS) as a function of  in-plane magnetic fields $B_\parallel$ and at different temperatures. In particular, we demonstrate that the gap of the topological surface state reduces with increasing in-plane field that rotates the magnetization towards the surface, thus decreasing the out-of-plane $M_\perp(T)$ (see Fig.~\ref{Fig1}a--c). Since the gap size  depends dominantly on $M_\perp$ \cite{Rosenberg12,Henk2012,Mong2010,Sekine2021}, this result  corroborates the magnetic origin of the gap. 
%The complex magnetization in the probed sample implies a preferential out-of-plane orientation of the magnetization vector ${\bf M}$ at $B_\parallel=0$\,T leading to an average dipole moment per Mn of $\approx 1.5\,\mu{\mathrm B}$ only  \cite{Wimmer2021}, i.e. significant disorder in the orientation of local Mn spins (Fig.~\ref{Fig1}a). 
While the applied in-plane field rotates the Mn spins preferentially  towards the in-plane direction, a large $B_\parallel\simeq 60$\,T is required for complete alignment (Fig.~\ref{Fig1}c) \cite{Wimmer2021,Lai2021}. This implies that a moderate $B_\parallel$  leads to a gap reduction with spatial fluctuations instead of a complete closure of the gap.
Indeed, we find that the magnetic gap reduces with  $B_\parallel$, but does not close completely up to $B_\parallel=3$\,T, where it exhibits significant spatial variation of the gap size. This is strong additional evidence that the origin of the gap is indeed the topological Dirac cone. Moreover, we find that the gap slowly fluctuates at elevated temperatures towards $T_{\rm C}$, where the remanent magnetization starts to decay. We map the spatiotemporal fluctuations by recording sequential gap size maps $\Delta(x,y)$ of the same area, featuring lateral patches of enhanced gap that slowly meander across the surface on the scale of hours, similar to magnetic domains close to a phase transition \cite{Shpyrko2007,Kronseder2015,Lachman16}.

Epitaxial \MnSbSbTe4\ films with 200 nm thickness were grown by molecular beam epitaxy \cite{Wimmer2021} and transferred in ultrahigh vacuum to two scanning tunneling microscopes (STMs) operating down to $T=4.3$\,K and $T=6$\,K, respectively \cite{Mashoff2009} (Supplementary Section S1).  The voltage $V$ is applied to the sample and the current $I$ is measured via the tip. The differential conductivity $dI/dV(V)$, proportional to the local density of states (LDOS) of the sample \cite{Morgenstern03}, is recorded by lock-in technique after opening the feedback loop at voltage $V_{\rm stab}$ and current $I_{\rm stab}$. 

Figures~\ref{Fig1}d and e sketch the atomic structure of the epitaxial \MnSbSbTe4\ as deduced by transmission electron microscopy, X-ray diffraction and Rutherford backscattering \cite{Wimmer2021}. Septuple layers with a Mn layer in the center are separated by van-der-Waals gaps between adjacent Te layers. Moreover, about 3.5\,\% of the atoms in the Sb layers are replaced by Mn atoms (green balls in the violet layers).  STM reveals the hexagonal structure of the top Te layer (Fig.~\ref{Fig1}f). The height fluctuations up to about 1\,\AA\ (RMS roughness: 24\,pm), have been attributed previously to the Mn atoms in the subsurface Sb layer \cite{Wimmer2021}. The $dI/dV$ curves exhibit a band gap at the Fermi level $E_{\rm F}$ for $B_\parallel=0$\,T (Fig.~\ref{Fig1}g, blue curve) \cite{Wimmer2021}. At the chosen fixed surface position, the band gap reduces in size at $B_\parallel=1$\,T (red curve) and completely disappears at $B_\parallel=3$\,T (yellow curve). This is in line with the expectation that, in first order, the gap size depends only on the magnetization perpendicular to the surface, while an in-plane component $M_\parallel$ only shifts the position in $k$ space but does not induce a gap \cite{Sekine2021,Rosenberg12,CXLiu08,Henk2012}. The reducing gap size, thus, indicates that the Mn spins are locally rotated into the plane by $B_\parallel$. 
\begin{figure}
\includegraphics*[scale=0.23]{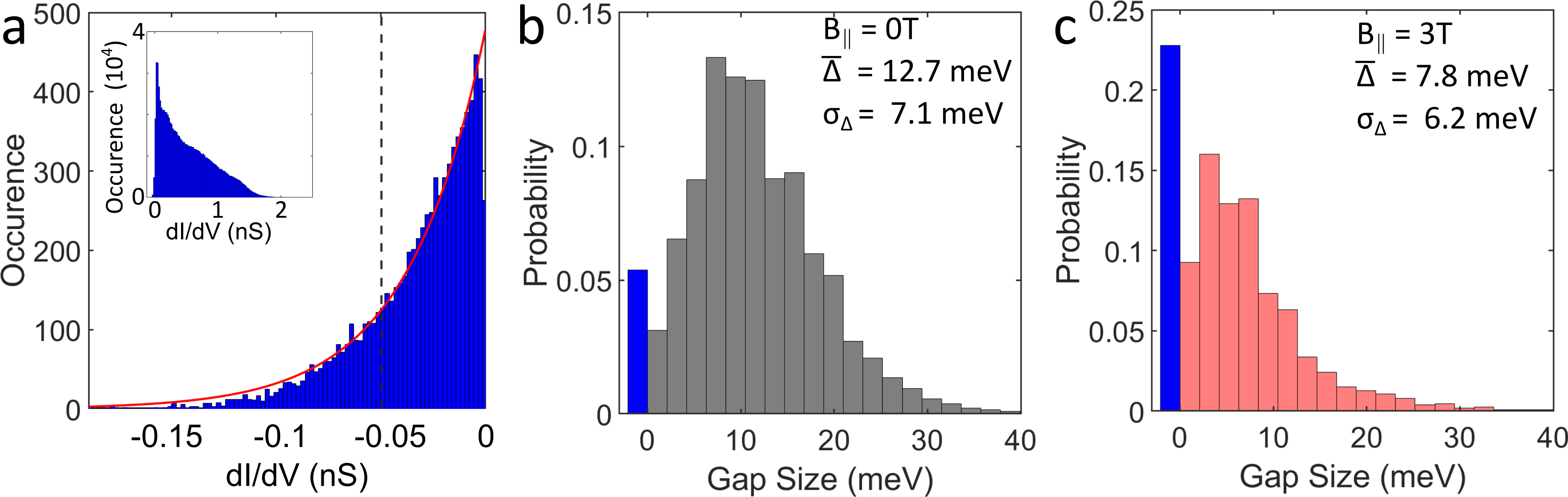}
\vspace{0cm}
\caption{{\bf}   (a) Histogram of negative $dI/dV$ values resulting from 2500 $dI/dV(V)$ curves  covering an area of 625\,nm$^2$, red line: exponential fit, black dashed line: threshold $dI/dV_{78 \%}$  used for gap determination (see text). Inset: full histogram including positive $dI/dV(V)$, 
$B_{\parallel}=0$\,T. (b) Histogram of deduced band gaps $\Delta$ at $B_{\parallel}=0$\,T for 5500 $dI/dV(V)$ curves covering 1375\,nm$^2$,  the blue bar at negative $\Delta$ counts all spectra without gap, while the bar covering $0-2$\,meV excludes these spectra, $\overline{\Delta}$: average gap value including the blue bar $\Delta =0$\,meV, $\sigma_\Delta$: standard deviation. (c) Same as b at $B_{\parallel}=3$\,T for 4600 $dI/dV(V)$ curves covering 1150\,nm$^2$. (a--c) $V=-50$\,mV to $50$\,mV, $V_{\rm stab}=-140$\,mV, $I_{\rm stab}=400$\,pA, $T=6$\,K. }
\label{Fig2}
\end{figure}

For quantitative band gap determination, we  must consider the noise level of $dI/dV(V)$ curves. Therefore, we firstly determine a histogram of all  $dI/dV <0$\,nS for a $dI/dV(V)$ map after careful calibration of the lock-in phase %using a complete $dI/dV(V)$ map
(Fig.~\ref{Fig2}a). We regard the negative $dI/dV$ as noise. The histogram can be fitted by an exponential function (red line, Fig.~\ref{Fig2}a). The $dI/dV$ value that encloses 78\,\% of the area below the exponential curve, $dI/dV_{78 \%}$ (dashed line, Fig.~\ref{Fig2}a), serves as threshold, i.e. only voltages $V$ with $dI/dV(V) < - dI/dV_{78 \%}$ are  regarded to belong to the gap (Supplementary Section S2). Figure~\ref{Fig2}b and c display the histograms of the resulting gap values $\Delta$ for $B_\parallel=0$\,T and 3\,T,  each using several $dI/dV(V)$ maps including $\sim 5000$ curves and covering an area as large as $\sim 1300$\,nm$^2$. The mean value $\overline{\Delta}$, that includes curves without gap as $\Delta =0$\,meV, decreases from $\overline{\Delta}(B_\parallel=0$\,T$)=12.7$\,meV to $\overline{\Delta}(B_\parallel=3$\,T$)=7.8$\,meV, i.e. 
the average gap reduces, 
%by applying $B_\parallel$, 
but does not close completely. The absolute gap values depend on the chosen threshold (Supplementary Section S2), but the threshold is identical for the two $B_\parallel$ and the gap reduction by $B_\parallel$ turns out to be robust for differently chosen thresholds. 
The percentage of curves without gap (blue bar at $\Delta <0$\,meV, Fig.~\ref{Fig2}b,c) increases from 5\,\% to 23\,\%. This indicates that only about 20\% of the Mn spins in the surface area are rotated in-plane, while the majority is only partially rotated towards the in-plane direction as evidenced by the reduced $\overline{\Delta}$. 
In-plane magnetization data of a ferromagnetic Mn$_{1.00}$Sb$_{2.09}$Te$_{3.91}$ bulk crystal revealed that $M_\parallel$ at $B_\parallel=3$\,T corresponds to $2\,\mu_{\mathrm B}$/Mn atom only, while $5\,\mu_{\mathrm B}$/Mn atom are achieved at $B_\parallel>60$\,T \cite{Lai2021}. This nicely corroborates the only partial rotation of Mn moments into the plane at $B_\parallel=3$\,T as in our case.  
Since time reversal symmetry is broken, a tiny  gap might also appear for a magnetization parallel to a topological insulator surface. However, according to density functional theory calculations of MnBi$_2$Te$_4$, this gap is an order of magnitude smaller than for surfaces with perpendicular magnetization \cite{Gu2021}.
The Dirac point energy $E_{\rm D}(x,y)$ (mid point energy of the gap) barely changes by applying an in-plane field ($B_\parallel=0$\,T: $\overline{E}_D=1.6$\,meV, $\sigma_{E_{\rm D}}=3.5$\,meV; $B_\parallel=3$\,T: $\overline{E}_D=1.2$\,meV, $\sigma_{E_{\rm D}}=4.0$\,meV, Supplementary Section S3). Hence, the change in local magnetization does not significantly influence the electrostatic potential. The fact that $\sigma_{E_{\rm D}}< \sigma_{\Delta}$ reveals that magnetic disorder is more severe than electrostatic potential fluctuations for MnSb$_2$Te$_4$. This is opposite for Cr-doped (Bi,Sb)$_2$Te$_3$ \cite{Chong2020}, where, however, spatial fluctuations of the magnetization vector $\bf{M}$ are still relevant for the magnetic field induced phase transition \cite{Lachman16,Liu2020}.

%\MM{Figure~\ref{Fig2}d--e display the histograms of the Dirac point energy $E_{\rm D}(x,y)$. The histograms are centered at $E_{\rm F}\simeq E_{\rm D}$ and are rather narrow ($\sigma_{E_{\rm D}}=??$\,meV). This is in contrast to Cr-doped Bi$_2$(Sb,Te)$_3$, where the gap distribution is narrow ($\overline{\Delta}\simeq 14$\,meV, $\sigma_\Delta \simeq 2$\,meV), but the $E_{\rm D}(x,y)$ fluctuates more strongly ($\sigma_{E_{\rm D}}=3.5$\,meV) \cite{Chong2020}. Hence, the substitutional doping by Cr or V leads to stronger and eventually limiting potential fluctuations, while the remaining substitutional doping by Mn leads to more severe spatial fluctuations of the magnetization. This fits to the observation that the limiting temperature for the QAHE is independent of $M(T)$ for Cr or V doped Bi$_2$(Sb,Te)$_3$ \cite{Figueroa2020,MogiAPL15,Xiao2018} but related to $M(T)$ for the Mn case \cite{DengNatPhys2020, Wimmer2021}.}  
%
\begin{figure}
\includegraphics*[scale=0.72]{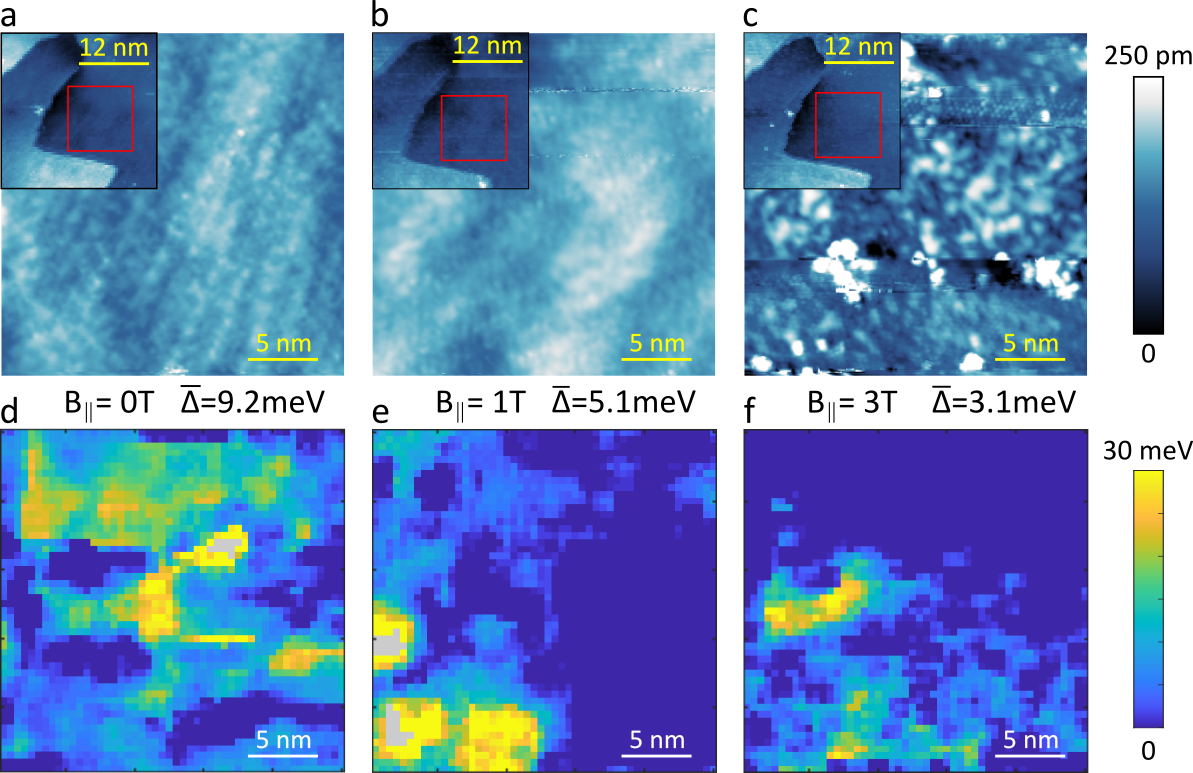}
\vspace{0cm}
\caption{{\bf}   (a)--(c) STM images of the same area, each recorded after 2D mapping of the $dI/dV(V)$ curves that are used to determine the $\Delta (x,y)$ map shown below. Insets: Larger scale STM images recorded prior to the $\Delta(x,y)$ maps, red rectangles: areas of main STM image and $\Delta(x,y)$ map below, $V=-500$\,mV, $I=250$\,pA. (d)--(f) $\Delta(x,y)$ maps of the identical area as displayed in the main image on top for the three $B_{\parallel}$ as marked, $V=-100$\,mV to $100$\,mV, $V_{\rm stab}=-100$\,mV, $I_{\rm stab}=400$\,pA, grey areas: gap can not be determined (see text), grey areas cover (d) 1.7\,\%,  (e) 3.4\,\%, (f) 0.7\,\%, (d) $\overline{\Delta} = 9.2$\,meV, $\sigma_\Delta = 7.2$\,meV, (e) $\overline{\Delta} = 5.1$\,meV, $\sigma_\Delta = 6.6$\,meV, (f) $\overline{\Delta} = 3.1$\,meV, $\sigma_\Delta = 5.5$\,meV. %Insets: correlation function $C(x)$ of the real space map (blue line) with exponential fit $C(x) \propto\exp{-x/\xi}$ (red line) and indicated correlation length $\xi$ (a--f) $T=6$\,K.
}
%Gray area relative sizes: 0T: 1.7\%, 1T: 3.4\%, 3T: 0.7\%, Gap sizes: 0T: $\Delta = 11.3 \pm 8.7$\,meV, 1T: $\Delta = 6.3 \pm 9.9$\,meV, 3T: $\Delta = 7.4 \pm 9.6$\,meV.

\label{Fig3}
\end{figure}
%

%Correlation length 0T: 2.3nm, 1T: 3.0nm, 3T: 3.0nm.
Figure~\ref{Fig3} shows $\Delta(x,y)$  maps of the identical area of \MnSbSbTe4\ for three different $B_\parallel$ values (Fig.~\ref{Fig3}d--f). The area is repeatedly adjusted by its relative position to step edges  as seen in the larger scale STM images (insets in Fig.~\ref{Fig3}a--c). Two experimental challenges appeared. Firstly, roughly 2 \% of the curves do not exhibit an increasing $dI/dV$ for $0{\,\rm mV}< V< 50$\,mV, such that $\Delta$ could not be determined \cite{Wimmer2021}. The corresponding positions are colored grey in the $\Delta (x,y)$ maps. The effect is repeatedly observed for several tips in both STM instruments up to $B_\parallel =3$\,T and up to $T=50$\,K. The reason is unknown, but might be related to dynamic trapping of charges from the tunneling current by defects. %(Supplementary Section \MM{S4}).
Secondly, the conditions for STS were not completely stable during the recording of $dI/dV(V)$ maps at $B_\parallel > 0$\,T as clusters of unknown origin appeared  on the surface after the maps (Fig.~\ref{Fig3}c). Such clusters can change the electrostatic potential and, hence, also $\Delta(x,y)$. Consequently, a reliable comparison can only be made between Fig.~\ref{Fig3}d and e, again demonstrating a reduction of the average gap size $\overline{\Delta}$ with the magnetic field applied, decreasing from $\overline{\Delta}$ = 9.2\,meV at $B_\parallel = 0$\,T to 5.1\,meV at $B_\parallel=1$\,T, in line with the result obtained from  the histograms (Fig.~\ref{Fig2}).
In the top part of the images, $\Delta(x,y)$  exhibits a rather homogeneous decrease of the local $\Delta$ with increasing $B_\parallel$. However, also regions  of local increases are found (lower left part) that encircle areas with undefined gap (grey regions), presumably indicating charge trapping.  The correlation length of $\Delta(x,y)$ slightly increases from $\xi_{6 {\rm K}}(0\,{\rm T})=2.1$\,nm to $\xi_{6 {\rm K}}(1\,{\rm T})=2.7$\,nm (Supplementary Section S3), being only slightly larger than the distance between individual Mn atoms within the subsurface Sb layer $d_{\rm Mn-Mn}\simeq 2$\,nm. %, i.e. the distance between 5 Te atoms along an atomic line. 
Hence, it appears that each Mn atom in the Sb layer contributes significantly to the local gap $\Delta(x,y)$. 
Figure~\ref{Fig3}f, recorded at $B_\parallel =3$\,T, must be regarded as a largely modified potential landscape due to the surface contamination found after recording $\Delta(x,y)$. Indeed, there is no obvious correlation to $\Delta(x,y)$ at lower $B_{\parallel}$. 
Nevertheless, the average gap of $\overline{\Delta}_{6 {\rm K}}(3\,{\rm T})=3.1$\,meV is further decreased by about 40\% with respect to $\overline{\Delta}_{6 {\rm K}}(1\,{\rm T})$. The average gap $\overline{\Delta}_{6 {\rm K}}(3\,{\rm T})$ in this area is relatively small (compare Fig.~\ref{Fig2}c), which is due to the large patch without gap observed in the upper part of the image.
The correlation length $\xi_{6 {\rm K}}(3\,{\rm T})=3.0$\,nm barely changes compared to $B_\parallel =1$\,T. Previously reported magnetization data   \cite{Wimmer2021} shows that the in-plane magnetization $M_\parallel$ of our \MnSbSbTe4\ films increases by about 20\,\% between $B_\parallel=1$\,T and $B_\parallel=3$\,T. Albeit a direct comparison is not straightforward, the changes of band gap and magnetization are of a very similar strength.
Hence, we conclude that the origin of reduced $\Delta(x,y)$ is dominated by a partial, non-uniform rotation of the Mn spins into the plane by  $B_\parallel$ as illustrated in Fig.~\ref{Fig1}b. Such a relation between gap and magnetization direction as expected for ferromagnetic TIs \cite{Tokura2019,Sekine2021} has never been demonstrated experimentally before.
\begin{figure}
\includegraphics*[scale=0.45]{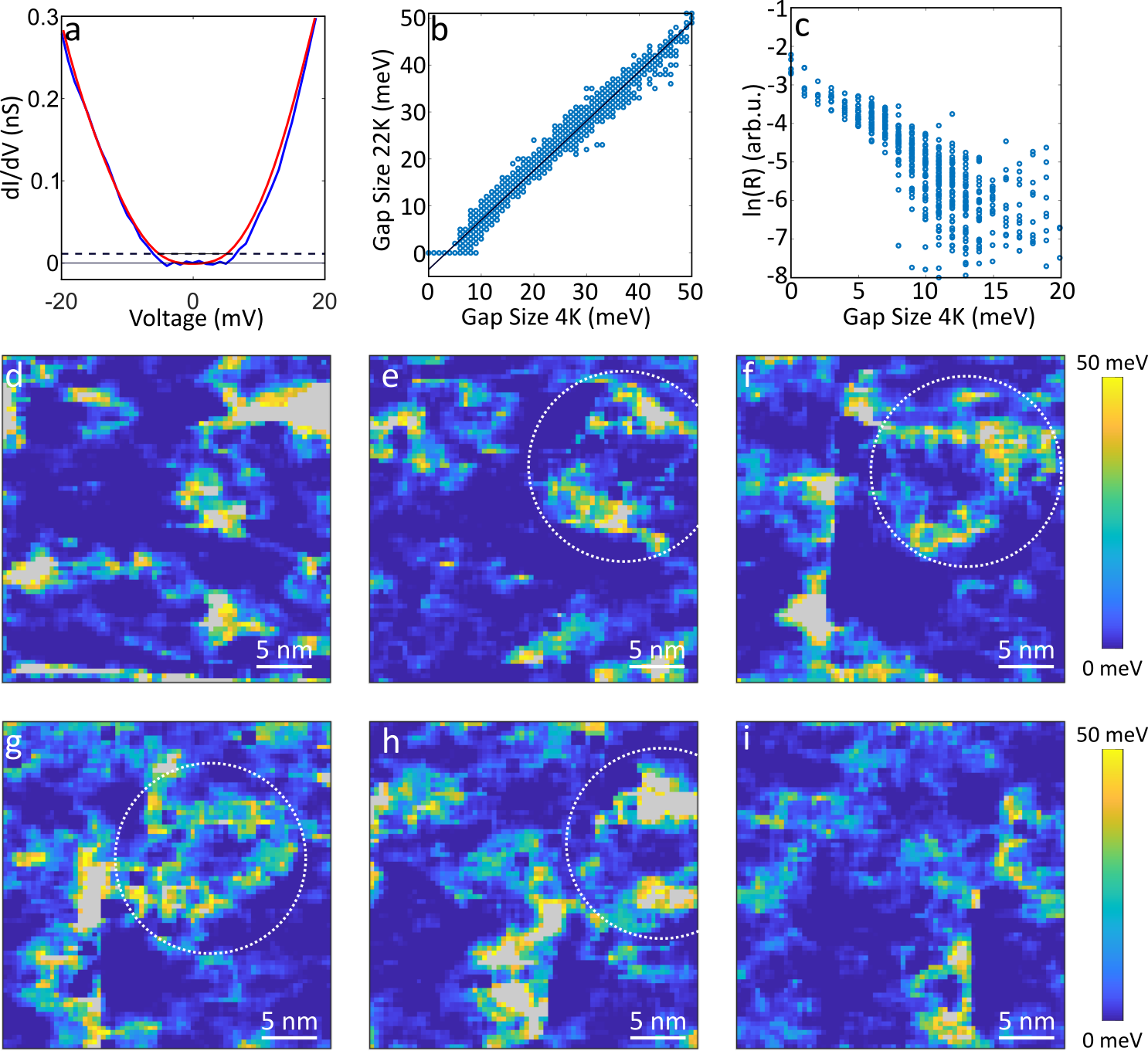}
\vspace{-0.6cm}
\caption{{\bf}   (a) Individual $dI/dV(V)$ curve recorded at $T=4.3$\,K (blue line), $V_{\rm stab}=-100$\,mV, $I_{\rm stab}=100$\,pA, $B_\parallel = 0$\,T. Red line: same  $dI/dV(V)$ after convolving with the Fermi-Dirac distribution $f(E,T=22\,{\rm K})$ according to eq.~\ref{eq:fold}. The dashed line indicates the threshold for gap determination ($-dI/dV_{95\,\%}$) deduced similar as in Fig~\ref{Fig2}a (Supplementary Section S2). The resulting gap sizes are $\Delta(4.3\,{\rm K})=13$\,meV,  $\Delta(22\,{\rm K})=10$\,meV.
(b) Scatter plot of the relation $\Delta_{\rm sim}(22\,{\rm K})$ to $\Delta(4.3\,{\rm K})$ for 7500 $dI/dV(V)$ curves recorded at $T=4.3$\,K.  
%covering an area of 625\,nm$^2$, 
$\Delta(4.3\,{\rm K})$ results from the recorded curves via the $-dI/dV_{95\,\%}$ threshold, $\Delta_{\rm sim}(22\,{\rm K})$ results after applying eq.~\ref{eq:fold}. Black line: linear fit. (c) Logarithmic plot of $R=dI/dV(0\,{\rm mV})/dI/dV(-50\,{\rm mV})$ as function of $\Delta(4.3\rm{\,K})$ for  7500 $dI/dV(V)$ curves, recorded at $T=4.3$\,K and folded with $f(E,22\,{\rm K})$ according to eq.~\ref{eq:fold} \cite{Wimmer2021}.
(d)--(i) Time series of subsequently recorded $\Delta(x,y)$ maps, $T=21-23$\,K continuously decreasing from d to i. Recording time per image: 4\,h.    
%white rectangles mark the identical area, 
Grey areas (2--6 \% of the images) indicate the positions where $\Delta$ could not be determined. The dashed circles in e--h mark a similar feature  (see text). Measurement conditions: $V=-100$\,mV to $100$\,mV, $V_{\rm stab}=-100$\,mV, $I_{\rm stab}=100$\,pA, $B_\parallel$=0\,T. 
}
\label{Fig4}
\end{figure}

Next, we probe spatiotemporal fluctuations of the magnetically induced gap $\Delta(x,y)$. $M_\perp(T)$ curves by XMCD \cite{Wimmer2021} revealed a continuous, mostly linear reduction of $M_\perp(T)$ for $T > 20$\,K . Consequently, we record $\Delta(x,y)$ maps slightly above, at $T=22\pm 1$\,K, in a series of images covering the same area (Fig.~\ref{Fig4}d--i). Indirect temperature stabilization led to a lateral drift of $0.2-2$\,nm/hour during the measurements (Supplementary Section S1). The determination of $\Delta$ at elevated $T$ must consider the Fermi level broadening of the probing tip  \cite{Wimmer2021,LeeIPNAS2015,SessiNC16,ChenNJP2015}. Figure~\ref{Fig4}a shows $dI/dV(V, T=4.3\,{\rm K})$ as recorded at $T=4.3$\,K (blue line). 
%The same curve after convolving with the derivative of the Fermi-Dirac distribution function $f(E,T)$, $dI/dV_{\rm sim}(V, 22\,{\rm K})$  (red line), 
Applying the convolution
\begin{equation}
\label{eq:fold}
\frac{dI}{dV}_{\rm sim}(V, T=22\,{\rm K}) :=\int_{-\infty}^{\infty}\frac{dI}{dV}\left(\frac{E}{e}, T=4.3\,{\rm K}\right) \cdot \frac{df(eV-E,T=22\,{\rm K}) }{dE}\hspace{2 mm} dE   
\end{equation}
with Fermi-Dirac distribution $f(E,T)=(e^{E/k_{\rm B}T}+1)^{-1}$ ($e$: electron charge, $k_{\rm B}$: Boltzmann constant, $E$: energy) returns the expected appearance of  $dI/dV(V)$ at 22\,K resulting from the same LDOS (red line). Evidently, the broadening due to the Fermi-Dirac distribution reduces the apparent $\Delta$ for identical LDOS. To account for this error, we employed a set of 7500 $dI/dV(V,T=4.3\,{\rm K})$ curves to determine the band gaps $\Delta(4.3 {\rm K})$ and applied eq.~\ref{eq:fold} to determine the apparent band gaps at 22 K, denoted $\Delta_{\rm sim}(22 {\rm K})$. The resulting scatter plot for the 7500 curves is shown in Fig.~\ref{Fig4}b with a linear fit (black line) of all data points exhibiting $\Delta_{\rm sim}(22 {\rm K}) \neq 0$\,meV. The fit is then used to relate measured $\Delta(22 {\rm K})$ to the expected $\Delta_{\rm exp}(4.3 {\rm K})$ of the same LDOS that we regard as the true gap of the LDOS at 22\,K: $\Delta_{\rm LDOS, 22\,{\rm K}}:=\Delta_{\rm exp}(4.3 {\rm K})$.  The method does not work for $\Delta(22 {\rm K})=0$\,meV, where we instead use   $R=dI/dV(0\,{\rm mV})/dI/dV(-50\,{\rm mV})$ to estimate $\Delta_{\rm LDOS, 22\,\rm K}$  (Fig.~\ref{Fig4}c). This more complex method has been established previously \cite{Wimmer2021} (Supplementary Section S2) and is relevant for 38\,\% of the curves leading to Fig.~\ref{Fig4}d--i. For $R>0.1$, one reliably gets  $\Delta_{\rm LDOS, 22\,\rm K}= 0$\,meV (Fig.~\ref{Fig4}c). 

Figure~\ref{Fig4}d--i show the resulting $\Delta_{\rm LDOS, 22\,\rm K}(x,y)$ maps in a time series. Here, $25-40$\,\% of the surface area exhibits $\Delta_{\rm LDOS, 22\,\rm K}=0$\,meV interrupted by irregularly shaped patches with $\Delta_{\rm LDOS, 22\,\rm K}>0$\,meV. The average gap size within the six images amounts to $\overline{\Delta}_{\rm LDOS, 22\,\rm K}=5-8$\,meV, while within the patches it is $\overline{\Delta}_{\rm LDOS, 22\,K, \Delta > 0}=8.5-10.5$\,meV. Both values are significantly smaller than the average gap size at $6$\,K (13\,meV) as expected from the fact that the magnetization decreases with increasing temperature \cite{Wimmer2021}.
The correlation length amounts to $\xi_{22 {\rm K}}=1.2-2.0$\,nm, i.e. slightly smaller than $\xi_{6 {\rm K}}=2.1$\,nm, but still close to $d_{\rm Mn-Mn}\simeq 2$\,nm (Supplementary Section S3). Most intriguingly, the patches change quasi-continuously in size and shape as a function of time. This is revealed by the dashed circles in Fig.~\ref{Fig4}e--h that identifies a similar structure that only slightly changes its shape. Since subsequent images differ by 4\,hours, the fluctuations are extremely slow. The same experiment performed at $T\simeq 25$\,K exhibits $\Delta(x,y)$ fluctuations as well, but due to the larger changes it was not possible to identify similar patches in subsequent images (Supplementary Section S3). All in all, spatiotemporal fluctuations of the gap size start at $T\simeq T_{\rm C}/2$, i.e. when the magnetization starts to decay. This important observation perfectly corroborates the magnetic origin of the gap. We note that $T=22$\,K is very close to the Neel temperature $T_{\rm N}=21$\,K of {\em antiferromagnetic} \MnSbSbTe4\ \cite{McQueeneyPRB19,ChenPRM20}. Hence, we conjecture that the competition between ferromagnetic and antiferromagnetic interlayer coupling driven by the local distribution of Mn atoms in the Sb layer \cite{liu2021,Riberolles2021,li2021,HuarXiv2020,Wimmer2021} is responsible for the magnetic fluctuations at this temperature well below $T_{\rm C}$. Recently, large $\bf{M}$ fluctuations well below $T_{\rm C}$ have also been found for Mn(Bi,Sb)$_{2n}$Te$_{3n+1}$ by ac magnetic susceptibility measurements \cite{Hu2021b} as well as for Cr-doped Sb$_2$Te$_3$ by spin resonance relaxation \cite{Figueroa2020}. This establishes $\bf{M}$ fluctuations as a general phenomenon in (Bi,Sb)$_2$Te$_3$ based magnetic topological insulators.   
We tried to quantify the spatiotemporal fluctuation at $T \simeq 22$\,K in more detail by various correlation functions, but could not establish a clear trend probably due to the limited data set and the remaining lateral drift during the measurements.

In summary, we have shown that the band gap of Mn-rich \MnSbSbTe4\ is controlled by the perpendicular magnetization of the surface. Thus, it is strongly reduced in size
by about 1/3 when an in-plane magnetic field of $B_\parallel =3$\,T is applied, which nicely agrees to the in-plane magnetization at this field being 1/3 of the saturation magnetization \cite{Lai2021}. At temperatures  approaching $T_{\rm C}$, the magnetic gap starts to fluctuate spatiotemporarily as measured for $T_{\rm C}/2$, in line with the starting reduction of remanent magnetization. Both results corroborate our assignment of the gap to a magnetic gap of the topological surface state. 
Moreover, our results establish  \MnSbSbTe4\ as a highly promising material for direct spatiotemporal investigation of  magnetization changes close to a quantum critical point. The criticality is likely induced by the competition between ferromagnetic and antiferromagnetic interlayer coupling that can be tuned, in principle, by $B$ field.

% \cite{Jiang2012,Kellner2017}.  other gaps in Topo mag \cite{LeeIPNAS2015,SessiNC16,ChenNJP2015}.  by larger-scale  potential fluctuations {\cite{BeidenkopfNP11,Pauly2015}.} 

\medskip
\textbf{Acknowledgements} \par %delete if not applicable))
We gratefully acknowledge insightful discussion with G.\,Bauer and  funding by the Deutsche Forschungsgemeinschaft via  Germany's Excellence Strategy --- Cluster of Excellence: Matter and Light for Quantum Computing (ML4Q) EXC 2004/1 --- 390534769, and by Mo 858/19-1, by the Graphene Flagship Core 3,  and by the Austrian Science Funds (Projects No. P30960-N27 and I3938-N27).
%, and the Impuls-und Vernetzungsfonds der Helmholtz-Gemeinschaft under grant No. HRSF-0067 (Helmholtz-Russia Joint Research Group). 
%\MM{the CzechNanoLab project LM2018110  funded by MEYS , the CEITEC Nano Research  Infrastructure, the Swedish Research Council (Project No. 821-2012-5144), the Swedish Foundation for Strategic Research (Project No. RIF14-0053),the Spanish Ministerio de Ciencia e Innovaci\'on (Project No. PID2019-103910GB-I00), Tomsk State University (Project No. 8.1.01.2018), and Saint Petersburg State University (Project No. 51126254)}. 

%\MM{Possible Reviewers: Bode, Davies, Sessi, L\"upke, Moodera, Yazdani, Andrei}
 \newpage

\section*{Supplementary Sections}
The Supplementary Sections describe the experimental and theoretical methods as well as additional data supporting the conclusions from the main text.

% References
\medskip

% Use the following code if you wish to generate your bibliography with BibTeX;
% replace the string "MSP-template" below with the name(s) of
% the BibTeX data base(s) you want to use.
% The resulting bibliography-output (the content of the .bbl file)
% must be pasted back into this file before submission.
% Please also include your BibTeX data base file(s) in your submission
%

\bibliography{Main.bbl}
 
% \vspace{0.25cm}
%\noindent{\bf Competing interests}. The authors declare no competing interests.

%\vspace{0.25cm}
%\noindent{\bf Data availability}. The data sets generated and analysed here are available from the corresponding authors on reasonable request.

\phantom{xxxx}
 
%\noindent {{\bf  Code   availability.}   The employed electronic structure codes   can   be downloaded after the corresponding licence requirements given on the  respective webpages are fulfilled.} % end rouge

%\vspace{0.25cm}
%\noindent{\bf Additional information}. Supplementary information is available for this paper at https://***
\medskip 
\phantom{xxxx}

\end{document}

\def\Tc{$T_{\rm C}$}
\def\TN{$T_{\rm N\'eel}$}
\def\Ef{$E_{\rm F}$}
\def\BiTe{Bi$_2$Te$_3$}
\def\BiSe{Bi$_2$Se$_3$}
\def\SbTe{Sb$_2$Te$_3$}
\def\BiSbTe{(Bi, Sb)$_2$Te$_3$}
\def\SbVTe{(Sb$_{1-x}$V$_x$)$_2$Te$_3$}
\def\BiMnTe{(Bi$_{1-x}$Mn$_x$)$_2$Te$_3$}
\def\BiMnSe{(Bi$_{1-x}$Mn$_x$)$_2$Se$_3$}
\def\MnSbSbTe4{MnSb$_2$Te$_4$}
\def\MnBiBiTe4{MnBi$_2$Te$_4$}
\def\MnBiBiSe4{MnBi$_2$Se$_4$}
\def\HgMnTe{Hg$_{1-x}$Mn$_x$Te}
\def\Gbar{$\overline{\Gamma}$}
\def\MnBi{Mn$_{\rm Bi}$}
\def\muB{$\mu_{\rm B}$}
\def\dIdV{${\rm d}I/{\rm d}V$}
\def\PK{\textcolor{blue}}
\def\MM{\textcolor{red}}
\def\OR{\textcolor{black}}
\def\bl{\textcolor{black}}
\def\ORa{\textcolor{black}}
\def\MMa{\textcolor{black}}
%\pagestyle{fancy}
%\rhead{\includegraphics[width=2.5cm]{vch-logo.png}}

\title{Supplementary Information:\\ Probing the magnetic band gap of the ferromagnetic topological insulator \MnSbSbTe4}

\author{P. K\"uppers$^{1}$, J. Zenner$^{1}$, S. Wimmer$^{2}$, G. Springholz$^{2}$, O. Rader$^{3}$, M. Liebmann$^1$, M. Morgenstern$^{1*}$}  % end authors

\affiliation{$^1$II. Institute of Physics B and JARA-FIT, RWTH Aachen University, 52074 Aachen, Germany}

\affiliation{$^2$Institut f\"ur Halbleiter- und Festk\"orperphysik, Johannes Kepler Universit\"at,
			Altenberger Stra\ss e 69, 4040 Linz, Austria}
			
\affiliation{$^3$Helmholtz-Zentrum Berlin f\"ur Materialien und Energie,
			 Albert-Einstein-Stra\ss e 15, 12489 Berlin, Germany}

\date{\today}

\maketitle

 \newpage
\newpage
\tableofcontents
\vspace{2 cm}
\noindent {Corresponding author: } 
M.~Morgenstern, email: mmorgens@physik.rwth-aachen.de %O. \newpage

%\section*{Supplementary Sections}
%The following sections describe the experimental and theoretical methods as well as additional data supporting the conclusions from the main text.

%\newpage
\renewcommand\thefigure{S\arabic{figure}} 
\renewcommand\thetable{S\Roman{table}}

\section{S1: Experimental details}
\label{sec:S1}
\subsection{Sample Growth}
\MnSbSbTe4\ films were grown by molecular beam epitaxy (MBE) on BaF$_2$(111) substrates using a Varian GEN II system. Compound \SbTe\  and  elemental Mn and Te sources were employed for control of stoichiometry and composition. Typical sample thicknesses were 200 nm. Deposition was carried out at a sample temperature of 290$^\circ$C at which perfect 2D growth is sustained independently of the Mn concentration, as verified by {\it in situ} reflection high energy electron diffraction and atomic force microscopy. The flux rates were calibrated by quartz microbalance measurements.

\subsection{Sample Characterization}
Samples were characterized by X-ray diffraction, Rutherford backscattering, transmission electron microscopy, magnetometry by superconducting quantum interference devices and X-ray magnetic circular dichroism as well as spin-polarized angular photoelectron spectroscopy. The results are described in detail elsewhere \cite{Wimmer2021} revealing that about 3.5\% of the atoms in the Sb layers are replaced by substitutional excess Mn atoms besides a Mn-Sb exchange of $\sim 10$\,\%, that the material is ferromagnetic with Curie temperature $T_{\rm C}\simeq 45$\,K, coercive field of about 200\,mT and out-of-plane anisotropy, as well as that the material is a topological insulator with a Dirac point of the topological surface state close to the Fermi level and that the topological surface state has an out-of-plane spin polarization close to the Dirac point that is parallel to the direction of magnetization.        
Density functional theory studies showed that the replacement of Sb by Mn is required to turn the material ferromagnetic, to achieve the large $T_{\rm C}$ and, via the resulting magnetic disorder, to open a bulk band gap in the otherwise Weyl type material.

\subsection{Scanning Tunneling Microscopy and Spectroscopy}
\label{sec:S9}
For the scanning tunneling microscopy (STM) experiments, the samples were transferred from the MBE system in Linz by an ultrahigh vacuum (UHV) suitcase to Aachen at pressure $p<10^{-9}$\,mbar without breaking UHV conditions.  
STM measurements were conducted in two home built UHV-STMs, one operating down to $T = 4.3$\,K without magnetic field and one operating down to $T=6$\,K with a 3D magnetic field \cite{Mashoff2009}. W or Cr tips were etched ex-situ and additionally prepared in UHV by field emission and voltage pulses on clean W(110) or clean Au(111) prior to inserting the MnSb$_2$Te$_4$ samples into the STM. Topography images were recorded in constant-current mode at a tunneling current $I$ and bias voltage $V$ applied to the sample. The \dIdV$(V)$ spectra were recorded after firstly stabilizing the tip-sample distance at voltage $V_{\rm stab}$ and current $I_{\rm stab}$. Afterwards, the feedback loop was opened and \dIdV\ was recorded using standard lock-in technique with modulation frequency $f = 1386$\,Hz and amplitude $V_{\rm mod} = 1.4$\,mV while ramping $V$. The spectra were normalized to account for remaining vibrational noise during stabilization by equilibrating the integral between $V_{\rm stab}$ and $V=0$\,mV. 
%$dI/dV$ is given in arbitrary units that must be multiplied by $3\cdot10^2$ to return the values in nS.
We crosschecked that $dI/dV$ curves barely depend on the chosen $I_{\rm stab}$  and, hence, on the tip-surface distance excluding a significant influence of tip induced band bending \cite{Morgenstern2000}.

For STM measurements at $T>20$\,K, the STM body was exposed to thermal radiation via opening the He radiation shield.  The LN$_2$ shield was also partially opened until a maximum $T \simeq 60$\,K was achieved and then closed again.
The subsequent cooling with open He shield considerably slowed down at $T\simeq 30$\,K such that recording of $dI/dV(V)$ maps became possible with lateral drifts in the range $0.2-2$\,nm/hour. Lowest temperature in this configuration was $T\simeq 20$\,K. 

\section{S2: Band Gap Determination}
\label{sec:S9b}
%
%\newpage
\subsection{Threshold Method}
\begin{figure}[h!]
%\vspace{-1cm}
\includegraphics*[scale=0.19]{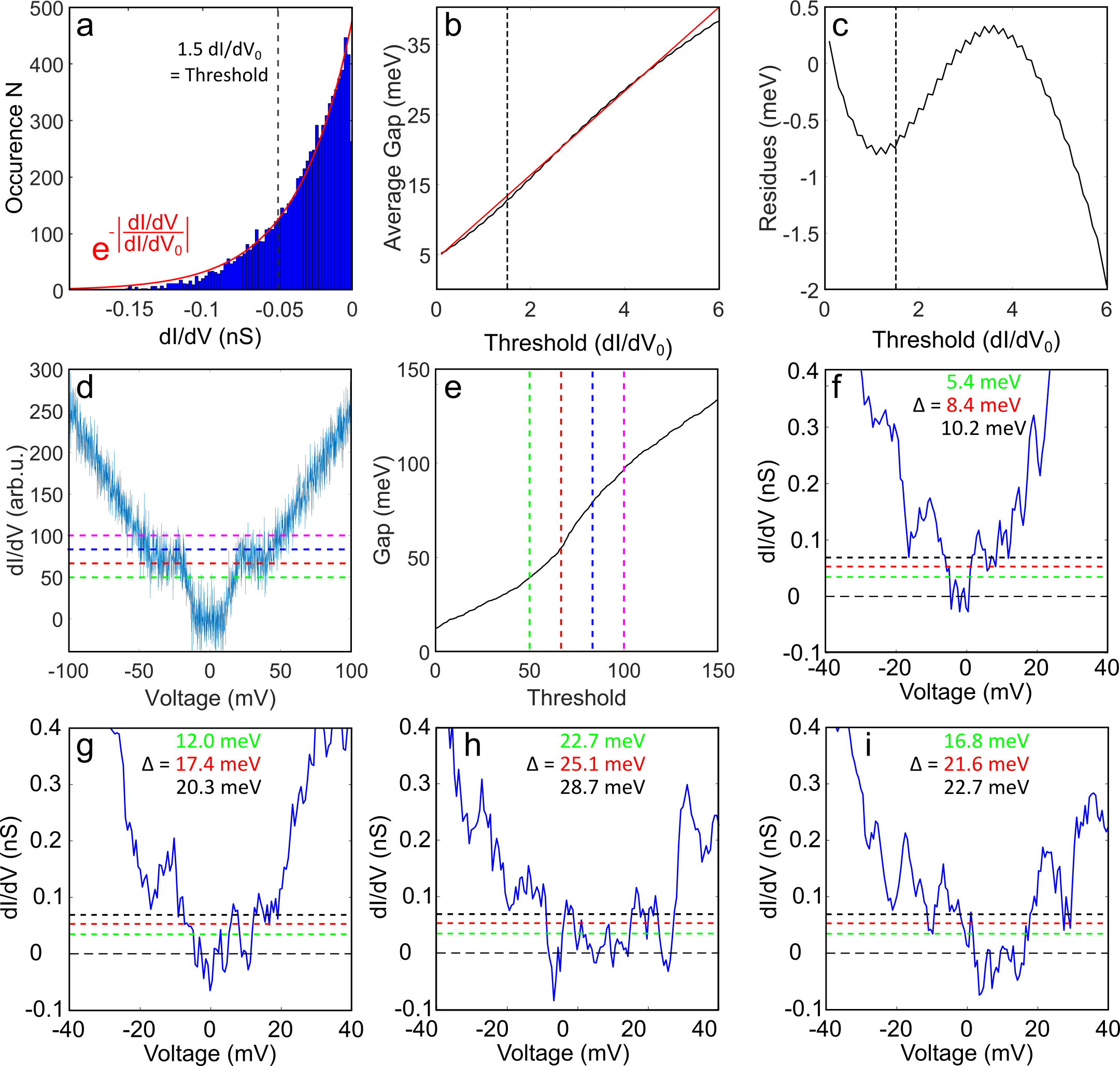}
\vspace{-0.4cm}
\caption{{\bf } 
(a) Histogram of negative $dI/dV$ values, 2500 $dI/dV(V)$ curves,  area: 625\,nm$^2$, red line: exponential fit $e^{-|dI/dV|/|dI/dV|_0}$, black dashed line: threshold as eventually used for the gap determination. %Same as Fig.~2a, main text.   
(b) Black line: average gap size $\overline{\Delta}$ as a function of the chosen threshold, red line: linear fit, dashed line: same threshold as in a, leading to $\overline{\Delta}=13$\,meV. (c) Residual of the linear fit in b, same dashed line as in a, b. (d) Artificial $dI/dV$ curve with gap and plateau surrounding the gap. A Gaussian noise of width $\sigma_{dI/dV}=18$~arb.~u. is added.
%, i.e. about half of the average $dI/dV$ at the plateau. 
Dashed lines: different thresholds for gap determination. (e) Determined gap size $\Delta$ from d as function of the threshold, emulating the curve $\overline{\Delta}({\rm threshold})$ from b, same colored dashed lines as in d. (f)--(i) Selected $dI/dV$ curves showing the problematic plateau close to the gap area with selected thresholds 
$1.0\cdot|dI/dV|_0$ (green line), $1.5\cdot|dI/dV|_0$ (red line), $2.0\cdot|dI/dV|_0$ (black line), resulting gaps $\Delta$ are indicated in the same color. (a)--(c), (f)--(i) 
%$V=-140$ to $100$\,mV, 
$V_{\rm stab}=-140$\,mV, $I_{\rm stab}=400$\,pA, $T=6$\,K.
}
\label{FigS1}
\end{figure}

As described in the main text, we employed the noise level  of the $dI/dV(V)$ curves to adequately determine the band gap region. The noise level was firstly reduced by averaging $3\times 3$ neighboring $dI/dV(V)$ curves covering an area of (1.5\,nm)$^2$.  Subsequently, an averaging of $dI/dV(V)$ across $\pm 1.2$\,mV along $V$ was employed. Then, we estimated the $dI/dV$ noise by plotting histograms of all negative $dI/dV$  for each $dI/dV(V)$ map (Fig.~\ref{FigS1}a). The exponential fit of such histograms $N(dI/dV) \propto \exp{-|dI/dV|/|dI/dV|_0}$ ($N$: occurence) is quite robust revealing  
a barely varying decay constant $|dI/dV|_0=0.0034\pm 0.0001$\,nS for several $dI/dV(V)$ maps at different $B_\parallel$ and $T$.
The choice of the threshold for gap determination $dI/dV_{x \%}$, enclosing $x$ \% of the area below the exponential, is not straightforward, since the apparent gap is partially surrounded by plateaus of relatively low $dI/dV$ (Fig.~\ref{FigS1}f--i).
Due to the observed contamination of the sample at smaller tip-sample distance, we had to operate at larger tip-surface distance implying a noise level that partially is as large as the surrounding plateaus. 
To deal with this difficulty, we estimated the average onset of the plateaus by plotting the average gap size $\overline{\Delta}$ as function of the chosen threshold (Fig.~\ref{FigS1}b). It turns out that this curve exhibits an S-shape (Fig.~\ref{FigS1}b--c) as expected for the plateau scenario (Fig.~\ref{FigS1}d--e).
The largest slope marks the position where $dI/dV_0$ matches the average $dI/dV$ of the plateaus. We used the residue of the S-shape from the linear fit (Fig.~\ref{FigS1}c) to find this point and have chosen the threshold slightly below to account for noise and variation in the plateau value of $dI/dV$. %to account for the apparent gap reduction by noise in the band gap region that overcomes the threshold 
Finally, we crosschecked the chosen threshold by manually inspecting the $dI/dV(V)$ curves with surrounding plateau (Fig.~\ref{FigS1}f--i). These curves are identified numerically by their relatively small slope of $dI/dV(V)$ in the area surrounding the determined gap. Examples are shown in Fig.~\ref{FigS1}f--i. The chosen threshold is the red dashed line. The selected threshold tends to slightly underestimate the gap by discarding areas within the gap because they exceed the threshold due to  noise. On the other hand, the black, higher  threshold  overestimates the gap because it adds areas of the surrounding plateaus. We omit to select the threshold more precisely, since we feel this is not justified. We stress that the resulting average low temperature gap at $B=0$\,T,
$\overline{\Delta}_{6\,{\rm K}}(B=0{\rm\,T})=13$\,meV, is rather close to the value determined previously at  lower $|V_{\rm stab}|$ and, hence, lower noise, $\overline{\Delta}_{4.3\,{\rm K}}(B=0{\rm\,T})=17$\,meV, but there with significantly less statistics \cite{Wimmer2021}.

The $dI/dV(V)$ maps in Fig.~4, main text, are also recorded at lower $|V_{\rm stab}|$ and, hence, exhibit less noise. Consequently, we could choose a larger threshold, $2.5\cdot |dI/dV|_0$, while still avoiding contributions from surrounding plateaus, and, hence, establish a more precise gap determination.

%\newpage

\subsection{Determination of Band Gaps by $dI/dV$ Relation} 
\label{sec:S9d}
 
 At larger $T$, band gaps $\Delta(x,y)$ in the LDOS of the sample can be completely masked by the Fermi level broadening of the tip. This typically applies for $k_{\rm B}T > \Delta/5$ \cite{Morgenstern03}, hence up to about $\Delta\simeq 10$\,meV
for $T=21$\,K. Hence, gaps with $\Delta < 10$\,meV would be determined as $\Delta =0$\,meV  by the threshold method (Fig.~4b, main text). 
Since the LDOS spatially fluctuates we can not deconvolve the $dI/dV$ spectra to recover the sample LDOS.  
In \cite{Wimmer2021}, we have developed an alternative method via the ratio $R=[\hbox{\dIdV}(0\hspace{1mm} {\rm mV})] / [\hbox{\dIdV}(-50\hspace{1mm} {\rm mV})]$ that turns out to be monotonously anticorrelated with the gap size $\Delta$. 
To show this, we used $dI/dV(V)$ curves recorded at 4.3\,K (8000 curves) and convoluted them with the derivative of the Fermi-Dirac distribution at 22\,K (as in Fig.~4a, main text) before $R$ is determined. Subsequently, each $R$ is related to its corresponding $\Delta(4.3\,{\rm K})$ (same $dI/dV(V)$ curve) as deduced via the threshold method. A scatter plot of the data is shown in Fig.~4c, main text.
We used the resulting median of the simulated $R$ values for each $\Delta(4.3\,{\rm K})$ to deduce $\Delta(T)$ from a measured $R(T)$ at elevated $T$. For gap sizes below 10\,meV, the error of the method is 2--4\,meV (variance of all $\Delta(4.3\,{\rm K})$ values belonging to the same $R$).
The reference voltage $V_{\rm ref}=-50$\,mV is chosen such that it is not influenced by temperature ($|{eV_{\rm ref}}|\gg 5 k_{\rm B}T$) or gap size ($|{eV_{\rm ref}}|\gg \max({\Delta})/2$) and not influenced by spectroscopic features that appear at larger $V$. We crosschecked that 
the exact value of $V_{\rm ref}$ barely influences the deduced $\Delta$ \cite{Wimmer2021}. 

\section{S3: Additonal Data from Evaluation of Band Gap Maps $\Delta(x,y)$}
\subsection{Histograms of Dirac Point Energy}
\begin{figure}
%\vspace{-1cm}
\includegraphics*[scale=0.25]{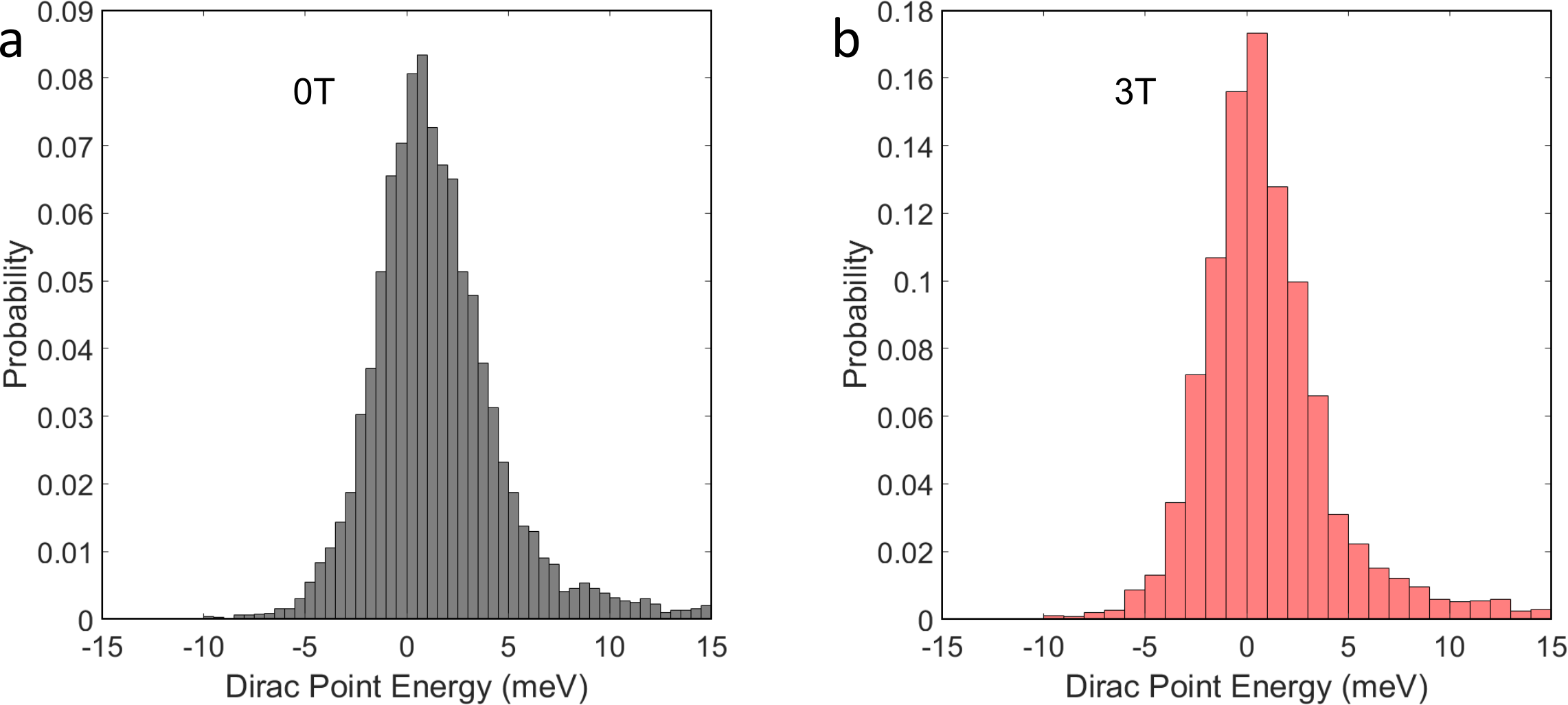}
%\vspace{-9.5cm}
\caption{{\bf } (a)--(b) 
Histograms of the Dirac point energy $E_{\rm D}$ (center of the determined band gap) at indicated in-plane $B$ fields $B_\parallel$. The same data set of $dI/dV(V)$ curves as in Fig.~2, main text, is evaluated, $T=6$\,K. Average values are $\overline{E}_{\rm D}(0\,{\rm T})=1.6$\,meV and $\overline{E}_{\rm D}(3\,{\rm T})=1.2$\,meV and standard deviations are $\sigma_{E_{\rm D}}(0\,{\rm T})=3.5$\,meV and $\sigma_{E_{\rm D}}(3\,{\rm T})=4.0$\,meV.}
\label{FigS1b}
\end{figure}

Figure~\ref{FigS1b} shows histograms of the center energies within the determined band gaps for two different $B_\parallel$. The center energies are regarded as Dirac point energy $E_{\rm D}$. The histograms exhibit mean values $\overline{E}_{\rm D}(0\,{\rm T})=1.6$\,meV and
$\overline{E}_{\rm D}(3\,{\rm T})=1.2$\,meV,
i.e. the band gap center favorably is very close to the Fermi level $E_{\rm F}$ as found previously for a distinct but identically prepared sample \cite{Wimmer2021}.
The standard deviation amounts to $\sigma_{E_{\rm D}}(0\,{\rm T})=3.5$\,meV and $\sigma_{E_{\rm D}}(3\,{\rm T})=4.0$\,meV. Hence, neither the average Dirac point energy nor its spatial fluctuation is significantly changing with $B_\parallel$ indicating that the electrostatic disorder barely depends on $B_\parallel$. The values of $\sigma_{E_{\rm D}}$ are rather similar to the values observed on Cr-doped (BiSb)$_2$Te$_3$
($\sigma_{E_{\rm D}}(0\,{\rm T})=3.3$\,meV), albeit in these samples $\overline{E}_{\rm D}(0\,{\rm T})\simeq 140$\,meV, such that the gap is not at $E_{\rm F}$, which generally reduces fluctuations of $E_{\rm D}$ by screening \cite{Chong2020}.

\subsection{Temporal Fluctuations at Larger Temperature}
\begin{figure}
%\vspace{-1cm}
\includegraphics*[scale=0.65]{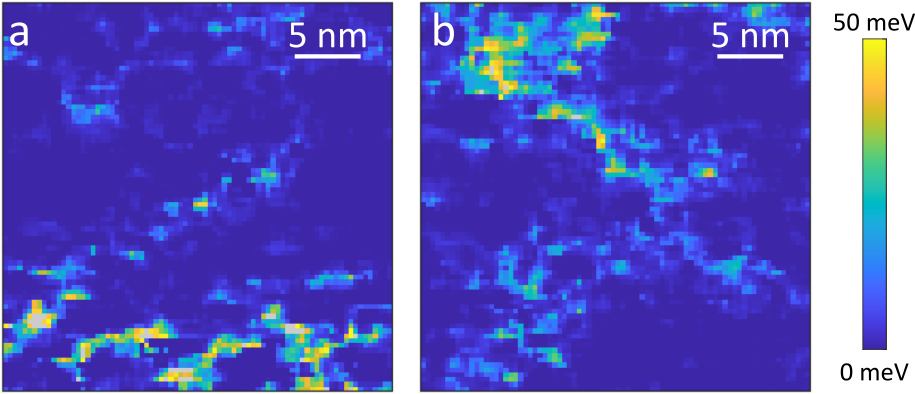}
%\vspace{-9.5cm}
\caption{{\bf } (a)--(b) 
Subsequently recorded $\Delta(x,y)$ maps,  $T=23-26$\,K continuously decreasing from a to b, recording time per image: 4\,h,
%white rectangles mark the identical area, 
grey areas ((a) 0.6 \% , (b) 0.1 \%) do not enable determination of $\Delta$, parameters for recording $dI/dV(V)$ curves: $V=-100$\,mV to $100$\,mV, $V_{\rm stab}=-100$\,mV, $I_{\rm stab}=100$\,pA, $B_\parallel$=0\,T.
%\PK{70x70 pixels, 30nm x 30nm}
}
\label{FigS2}
\end{figure}

Figure~\ref{FigS2} shows two $\Delta(x,y)$ maps recorded directly after each other at $T\simeq25$\,K (same way of $\Delta$ determination as in Fig.~4, main text). The recording of each image lasts 4 hours. Here, about 40-50\,\% of the probed area exhibit $\Delta=0$\,meV, which is slightly larger than at $T\simeq 22$\,K (25-40\,\%). The correlation length is barely changed (25\,K: $1.3-2.1$\,nm, 21\,K: $1.2-2.0$\,nm). However, it gets largely impossible to identify the temporal development of patches with finite $\Delta(x,y)$, indicating a faster fluctuation of magnetization and consequently of $\Delta(x,y)$. This is in line with expectations at larger $T$.    

\subsection{Spatial Correlation Functions}
\begin{figure}
%\vspace{-1cm}
\includegraphics*[scale=0.35]{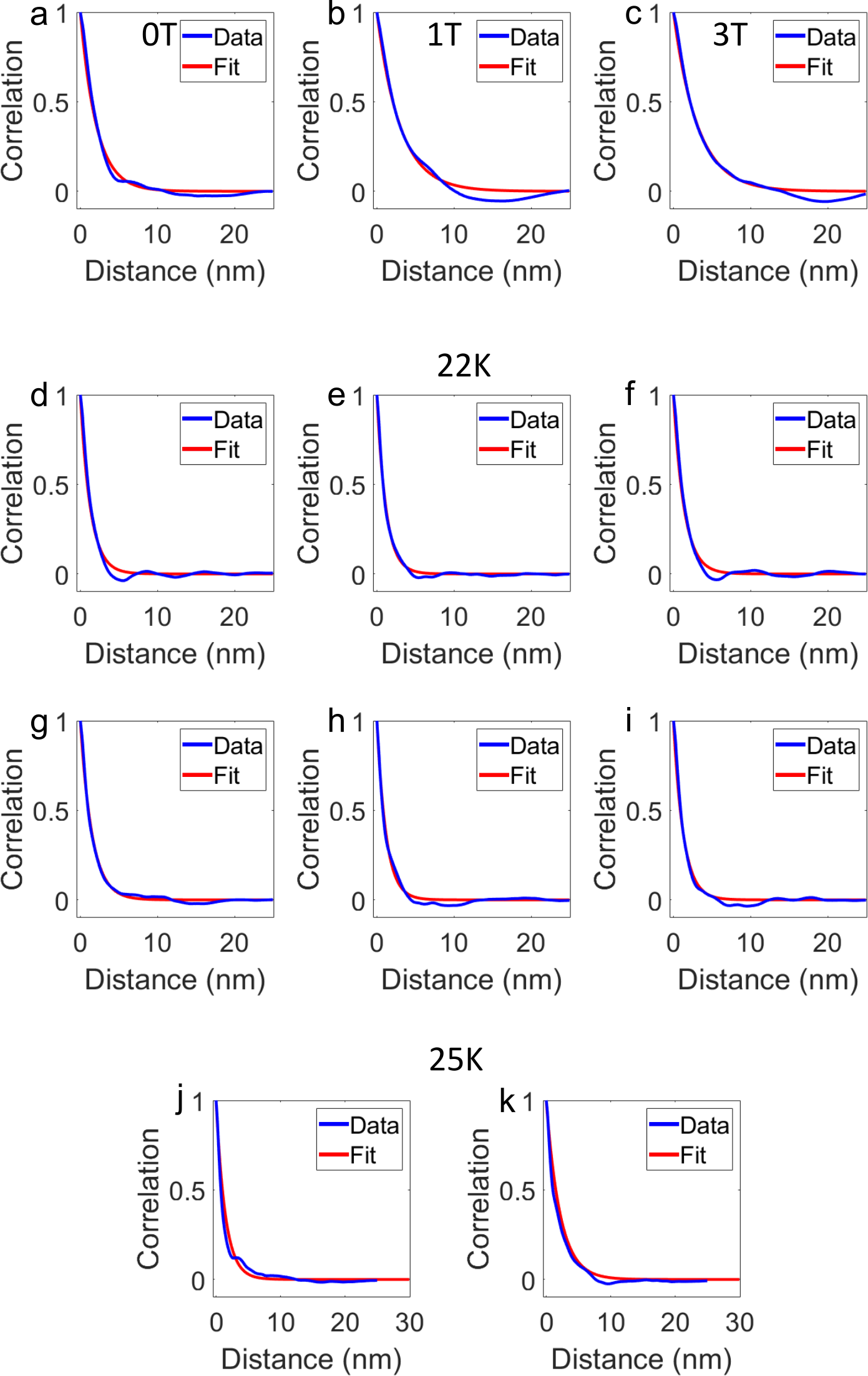}
%\vspace{-9.5cm}
\caption{{\bf } 
(a)--(c) Correlation functions $S(\Delta r)$ (blue) and exponential fits (red) of the $\Delta(x,y)$ maps of Fig.~3d--f, main text, respectively, $T=6$\,K, $B_\parallel$ as marked. (d)--(i) same as a, but for the $\Delta(x,y)$ maps of Fig.~4d--i, main text, respectively, $T\simeq 22$\,K, $B_\parallel=0$\,T. (j)--(k)
same as a, but for the $\Delta(x,y)$ maps of Fig.~\ref{FigS2}a--b, respectively,  $T\simeq 25$\,K, $B_\parallel=0$\,T.
}
\label{FigS3}
\end{figure}

Figure~\ref{FigS3} shows the correlation functions $S(\Delta r)$ (blue lines, $\Delta r$: distance between recording points) of the various $\Delta(x,y)$ maps at different $T$ and $B_\parallel$ as marked. Exponential fits $S_{\rm fit}(\Delta r)=\exp{\left(-\Delta r/\xi\right)}$ are added (red lines). Generally, the exponential decay fits well up to about $2 \xi$ and deviates at larger distances, where oscillations around  
$S(\Delta r)=0$ are partially observed. While at $T=6$\,K only the first negative half cycle of a possible oscillation is visible, the correlation functions at $T=22$\,K consistently reveal oscillations with a wavelength of about 10\,nm.  This indicates repulsive interactions between ferromagnetic puddles that might be investigated in more detail in future studies. The $S(\Delta r)$ curves at $T=25$\,K are not conclusive yet. 

Finally, we summarize the deduced data from the $\Delta(x,y)$ maps presented in Fig.~3d--f, main text, Fig.~4d--i, main text, and Fig.~\ref{FigS2}.
The average gap $\overline{\Delta}$ and the standard deviation of the $\Delta$ histogram $\sigma_{\Delta}$ are shown. Moreover, we evaluate the area that does not show a gap $A_{\Delta=0}$, the average gap in the remaining area
$\overline{\Delta}_{\Delta >0}$ and the correlation length $\xi$ (Fig.~\ref{FigS3}). Finally, the area where a gap could not be determined due to a flat curve with $dI/dV \simeq 0$\,nS up to the largest $V$ is given, showing a slight decrease at highest temperature. This might indicate that these curves are related to a current induced charge trap that is regularly decharged by thermal excitations at large enough $T$.     

\begin{table}[h]
\caption{Evaluation of all displayed $\Delta(x,y)$ maps recorded at temperature $T$ and in-plane magnetic field $B_\parallel$: $\overline{\Delta}$: average gap size, $\sigma_{\Delta}$: variance of gap sizes, $\overline{\Delta}_{\Delta >0}$: average gap size excluding the areas with $\Delta(x,y)=0$, $A_{\Delta=0}$: relative area with $\Delta(x,y)=0$, $\xi$: correlation length from exponential fit of correlation function (Fig.~\ref{FigS3}), $A_{\rm excluded}$: relative area of $dI/dV(V)$ curves that do not raise above the threshold at positive $V$ up to the maximum $V$.}
\vspace{0.5cm}
\begin{tabular}{|c|c|c|c|c|c|c|c|c|}
\hline
Figure & $T$ (K) & $B_\parallel$ (T)  & $\overline{\Delta}$ (meV) & $\sigma_{\Delta}$ (meV)& $\overline{\Delta}_{\Delta >0}$ (meV)  & $A_{\Delta=0}$ (\%) & $\xi$ (nm) & $A_{\rm excluded}$ (\%)\\ \hline
3d  & 6  & 0             & 9.2      & 7.2     & 11.0                       & 16                    & 2.1   & 1.5                    \\ \hline
3e & 6  & 1             & 5.1     & 6.6     & 10.3                        & 51                    & 2.7  & 3.7                    \\ \hline
3f & 6  & 3             & 3.1      & 5.5     & 7.1                         & 57                    & 2.5  & 0.02                   \\ \hline
4d & 23  & 0             & 5.7      & 10.8     & 10.0                        & 40                    & 1.5   & 6.0                   \\ \hline
4e & 23  & 0             & 5.1      & 9.8      & 8.4                        & 38                    & 1.4  & 1.7                     \\ \hline
4f & 22  & 0             & 6.4      & 10.4     & 9.8                        & 34                    & 1.4   & 2.8                   \\ \hline
4g & 22  & 0             & 7.9      & 11.1     & 10.5                        & 24                    & 1.6   & 2.5                   \\ \hline
4h & 21  & 0            & 5.1      & 11.0     & 10.0                        & 28                    & 2.0   & 5.2                   \\ \hline
4i & 21  & 0             & 7.5      & 8.7      & 7.4                        & 30                    & 1.2   & 0.6                      \\ \hline
S2a & 25 & 0      & 3.1      & 7.6      & 6.3                      & 51                    & 1.3   & 0.6                      \\ \hline
S2b & 24 & 0     & 4.1      & 7.6      & 6.6                        & 38                    & 2.1   & 0.1                    \\ \hline
\end{tabular}
\end{table}
%Correlation length 22K of average Correlation function: 1.37nm

%\begin{table}[]
%\begin{tabular}{|l|l|l|l|l|l|l|}
%\hline
%Image Number & $\Delta$ & $\sigma$ & $\Delta$ without 0 & Percentage Gray & Percentage $\Delta=0$ & Correlation length (nm) \\ \hline
%\end{tabular}
%\end{table}

% References
\medskip

% Use the following code if you wish to generate your bibliography with BibTeX;
% replace the string "MSP-template" below with the name(s) of
% the BibTeX data base(s) you want to use.
% The resulting bibliography-output (the content of the .bbl file)
% must be pasted back into this file before submission.
% Please also include your BibTeX data base file(s) in your submission
%

\bibliography{Supplement.bbl}

% \vspace{0.25cm}
%\noindent{\bf Competing interests}. The authors declare no competing interests.

%\vspace{0.25cm}
%\noindent{\bf Data availability}. The data sets generated and analysed here are available from the corresponding authors on reasonable request.

\phantom{xxxx}
 
%\noindent {{\bf  Code   availability.}   The employed electronic structure codes   can   be downloaded after the corresponding licence requirements given on the  respective webpages are fulfilled.} % end rouge

%\vspace{0.25cm}
%\noindent{\bf Additional information}. Supplementary information is available for this paper at https://***
\medskip 
\phantom{xxxx}